\documentclass[prd, aps, twocolumn, nofootinbib, showpacs, superscriptaddress,longbibliography]{revtex4-2}

\usepackage[english]{babel}
\usepackage[utf8]{inputenc}
\usepackage{bm}
\usepackage{subfigure}
\usepackage{amsmath}
\usepackage[T1]{fontenc}
\usepackage{esint}
\usepackage{twoopt}
\usepackage{float}
\newcommand{\fig}[1]{Fig.~\ref{#1}}

\newcommand{\svu}[2]{\mathbb{#1}^\mathfrak{#2}}
\newcommand{\svl}[2]{\mathbb{#1}_\mathfrak{#2}}
\newcommand{\jmat}[2]{{\svu{J}{#1}}_\mathfrak{#2}}
\newcommand{\lmat}[2]{{\svu{L}{#1}}_\mathfrak{#2}}
\newcommand{\vt}{\vartheta}
\newcommand{\veff}{V_{\rm eff}}

\usepackage{amsmath, amssymb, amsthm, amsfonts}
\usepackage{mathtools}
\usepackage{physics}
\usepackage{xcolor}
\usepackage{graphicx}
\usepackage{adjustbox}
\usepackage{placeins}
\usepackage[T1]{fontenc}
\usepackage{lipsum}
\usepackage{csquotes}
\usepackage[colorlinks=true,citecolor=green,urlcolor=blue,linkcolor=blue,linktocpage, breaklinks, pdftex, pdftitle={Article}, pdfauthor={Author}]{hyperref}

\usepackage[utf8]{inputenc}
\usepackage{latexsym}
\usepackage{rotating}

\begin{document}
\title{Lyapunov Exponents to Test General Relativity}

\author{Alexander Deich}
    \email[Correspondence email address: ]{adeich2@illinois.edu}
    \affiliation{Illinois Center for Advanced Studies of the Universe, Department of Physics, University of Illinois at Urbana-Champaign, Champaign, Illinois, USA}

\author{Nicol\'as Yunes}
\affiliation{Illinois Center for Advanced Studies of the Universe, Department of Physics, University of Illinois at Urbana-Champaign, Champaign, Illinois, USA}

\author{Charles Gammie}
\affiliation{Illinois Center for Advanced Studies of the Universe, Department of Physics, University of Illinois at Urbana-Champaign, Champaign, Illinois, USA}
\affiliation{Department of Astronomy, University of Illinois at Urbana-Champaign}
\affiliation{NCSA, University of Illinois at Urbana-Champaign}

\date{\today}
\begin{abstract}
Photon rings are key targets for near-future space-based very-long baseline interferometry missions. The ratio of flux measured between successive light-rings is characterized by the Lyapunov exponents of the corresponding nearly-bound null geodesics. Therefore, understanding Lyapunov exponents in this environment is of crucial importance to understanding black hole observations in general, and in particular, they may offer a route for constraining modified theories of gravity. While recent work has made significant progress in describing these geodesics for Kerr, a theory-agnostic description is complicated by the fact that Lyapunov exponents are time-parameterization dependent, which necessitates care when comparing these exponents in two different theories. In this work, we present a robust numerical framework for computing and comparing the Lyapunov exponents of null geodesics in Kerr with those in an arbitrary modified theory. We then present results obtained from calculating the Lyapunov exponents for null geodesics in two particular effective theories, scalar Gauss-Bonnet gravity and dynamical Chern-Simons gravity. Using this framework, we determine accuracy lower-bounds required before a very-long baseline interferometry observation can constrain these theories.

\end{abstract}

\maketitle

\section{Introduction}
The Event Horizon Telescope (EHT) and its space-based successors will soon offer unprecedented views of the structure of photon trajectories in the space very near the photon ring of a black hole~\cite{Johnson_2023,akiyama2019first}.  These trajectories can be characterized by the number of fractional orbits the photon completes before scattering to the detector, giving rise to the so-called ``sub-rings''. The ratio of flux received between adjacent sub-rings is determined by the Lyapunov exponent at the photon ring. These Lyapunov exponents are therefore an observable quantity, and offer a route to understand the phase space of the environment they are observed in.  As measures of the stability of phase space trajectories, Lyapunov exponents open a window to the underlying physics of any given system. In particular, they provide clues to the underlying theory of gravity in play, and may therefore offer a way to use very-long baseline interferometry (VLBI) observations to place constraints on modified theories~\cite{johnson_universal_2020,gralla_lensing_2020}.

Impressive strides have been made in understanding the structure of null orbits near black holes described by the Kerr metric~\cite{gralla_lensing_2020,Gralla_2020}. This includes full analytic evolution equations for null geodesics, which have in turn enabled a much deeper understanding of the structure of the photon ring that a VLBI instrument could detect.  As a result, we now also have full analytic expressions for the Lyapunov exponents associated with the photon ring around a Kerr black hole. As Lyapunov exponents provide a method of testing the Kerr hypothesis, one may wish to also calculate Lyapunov exponents in different theories of gravity and investigate whether any difference from the Kerr result is indeed detectable.

Black holes in modified gravity, however, are not necessarily described by the Kerr metric. For example, two theories, scalar-Gauss-Bonnet (sGB)~\cite{Kanti:1995vq} and dynamical-Chern-Simons (dCS) gravity~\cite{dcs2003}, predict distinct modifications to the spacetime of rotating black holes, and therefore, to the particular trajectories that geodesics follow in each theory. Such modifications will necessarily also affect the values that Lyapunov exponents take in these theories.  Exact, analytic forms of these metrics remain unknown, leaving us with only approximate analytic forms written as power series in the black hole spin. Recent work, nonetheless, allows us to calculate these expansions to essentially arbitrary order in  spin~\cite{Cano_2019}.

The calculation of Lyapunov exponents in GR is simplified by certain symmetries that are absent in modified gravity theories. This machinery relies on the existence of two Killing vectors (associated with stationarity and axisymmetry) and one Killing tensor (associated with a Carter constant) that render the geodesic equations separable and integrable~\cite{gralla_lensing_2020,staelens2023black}. This is why Lyapunov exponents can be calculated analytically for the Kerr metric, a simple, parametric deformations of it~\cite{Johannsen_2013}, and the Manko-Novikov metric~\cite{Manko_Novikov_1992,manko_novikov_2011}. Separability and integrability, however, is far from guaranteed in modified gravity, black hole spacetimes. In particular, the nonexistence of a Carter-like constant for the spin-expanded sGB and dCS metrics beyond leading order in spin~\cite{Deich_2022,owen2021petrov} prevents us from casting the geodesic equations in terms of elliptic integrals, and thus, from calculating the Lyapunov exponents analytically.  

Therefore, a clear need exists to calculate Lyapunov exponents accurately and in a theory-agnostic way; it is to this end that we present the current work.  In this paper, we introduce a framework to calculate equatorial Lyapunov exponents accurately for arbitrary, axisymmetric modifications to the Kerr metric, allowing us to predict the flux ratio between adjacent sub-rings, and in turn, to potentially test the Kerr hypothesis with VLBI images.  We also develop a robust numerical method to check the calculation to high accuracy. We then implement this method on the two example theories described above, sGB and dCS gravity. We find the corrections are of ${\cal{O}}(10^{-1})$ for geodesics around dCS black holes, and ${\cal{O}}(10^{-0.5})$ for geodesics around sGB black holes. For any constraint to be realized, this method would also have to be used in conjunction with independent measurements of the black hole's spin and mass, as both of these also affect the size of the Lyapunov exponent.

The remainder of this paper is organized as follows. In Sec.~\ref{sec:le_def}, we give a primer on symplectic geometry and provide a full derivation of Lyapunov exponents in general relativity.  Then, in Sec.~\ref{sec:photonring}, we describe the structure of the photon shell, photon ring, and its subrings in general relativity. We discuss in Sec.~\ref{sec:PRlyaps} how Lyapunov exponents are calculated for the photon rings of Kerr black holes.  In Sec.~\ref{sec:modgrav}, we briefly describe the modified theories of gravity under consideration and the black hole solutions permitted by them. Then, we cover how we calculate Lyapunov exponents for black holes in these theories, before presenting our results for dCS and sGB gravity. Finally, in Sec.~\ref{sec:conclusions} we conclude and point to future research. Appendix~\ref{ap:lyap_pert} discusses the application of eigenvalue perturbation to Lyapunov exponents.  Throughout this paper, we use the convention of $G=1=c$.


\section{Symplectic Geometry and Lyapunov Exponents}\label{sec:le_def}

In this section, we go over the basics of symplectic geometry, which can be thought of as a theoretical prerequisite to formally understand Lyapunov exponents. We then develop the theory of Lyapunov exponents thoroughly, and briefly discuss several subtle details, which will be of importance for the present work.

\subsection{Phase Space Evolution and Symplectic Geometry}

The study of Lyapunov exponents requires manipulating objects that are constructed out of both positions and momenta.  We therefore must be careful: these constituents transform differently when contracted with a metric tensor, and so any object we construct from them will transform differently from more familiar objects.  The primary object of focus will be a set, $\svu{X}{}$, of components of coordinates, $q^\mu$, \emph{and} momenta, $p_\mu$, which live in an $n-$dimensional phase space. The set $\svu{X}{}$ therefore has $2n$ elements, and we index these with a gothic letter.  Explicitly,
\begin{align}
    \svu{X}{a} &= q^\mu, \qquad \mathfrak{a}\in [0,...,n-1],\\
    \svu{X}{a} &= p_{\mu-n}, \qquad \mathfrak{a}\in [n,...,2n-1].
\end{align}
The evolution of $\svu{X}{a}$ is governed by Hamilton's equations of motion, which in this context take the form
\begin{equation}
\label{eq:evolution}
    \svu{\dot{X}}{a} = \Omega^{\mathfrak{ab}}\svl{\partial}{b}H,
\end{equation}
for a Hamiltonian $H$, where $\svl{\partial}{b} = \partial/\partial\svu{X}{b}$, where the dot refers to a derivative with respect to proper time and where we have introduced the \emph{symplectic matrix}, $\Omega^{\mathfrak{ab}}$. This matrix is a $2n \times 2n$ matrix whose elements take the form~\cite{jose_classical_1998}
\begin{equation}
\Omega^{\mathfrak{ab}} = \begin{pmatrix}
    0_n & \mathbb{I}_n \\
   -\mathbb{I}_n & 0_n
\end{pmatrix},
\end{equation}
where $\mathbb{I}_n$ and $0_n$ are the $n \times n$ identity and null matrices, respectively.

\subsection{Analytic derivation of Lyapunov exponents}
\label{sec:deriv}

\begin{figure*}
    \centering
    \includegraphics[scale=0.35]{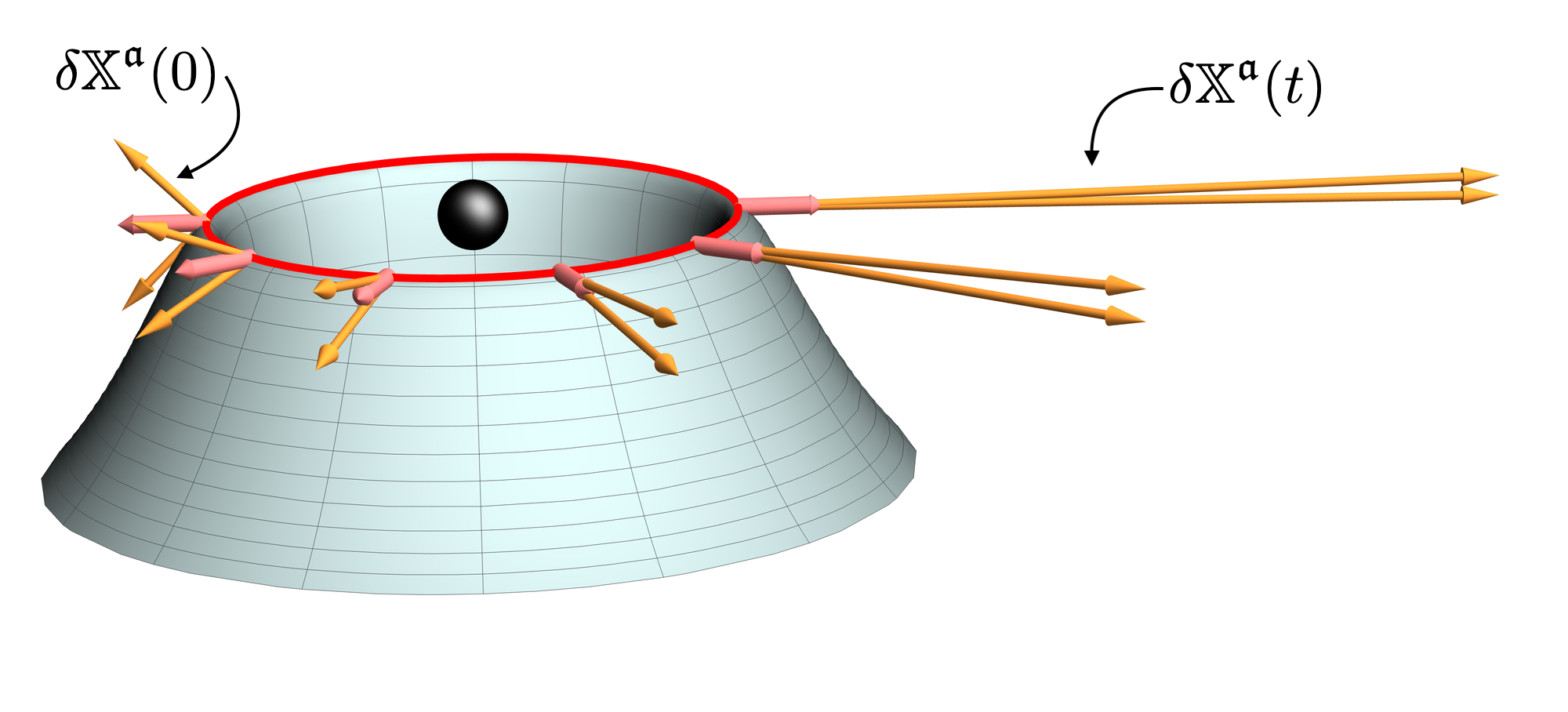}
    \caption{A representation of a photon ring orbit (red line) around a Schwarzschild black hole (black sphere), with the effective potential (blue surface) visualized.  Two initially orthogonal vectors ($\delta \svu{X}{a}(0)$, orange arrows furthest on the left) tangent to the phase space evolve under the linear stability matrix.  As they evolve counter-clockwise when looking at the black hole from above, they are rotated to align with the biggest eigenvector of the stability matrix (pink arrows). The log of the rate at which the magnitude of $\delta \svu{X}{a}(t)$ grows is the principal Lyapunov exponent.  In this cartoon, the vectors are drawn with respect to a vertical $p_r$ axis, while the height of the effective potential has units of energy. The direction pointing radially out from the black hole is the Schwarzschild $r$ coordinate.}

    \label{fig:vectorplot}
\end{figure*}

Lyapunov exponents are motivated by asking how a given trajectory responds to small perturbations in its initial phase space conditions.  Given some phase space trajectory $\svu{X}{a}(t)$ whose evolution is governed by a Hamiltonian $H$, as in Eq.~\eqref{eq:evolution}, we can look at how a perturbation to 
$\svu{X}{a}(t)=\svu{X}{a}{}^{(0)}(t)+\delta \svu{X}{a} (t)$
evolves by linearizing 
$\svu{\dot{X}}{a}(t)$ about small 
$\delta\svu{X}{a}(t)$:
\begin{align}\label{eq:deltadef}
    \delta \svu{\dot{X}}{a}(t)&=\jmat{a}{b}(t) \delta \svu{X}{b}(t),\\ \jmat{a}{b}(t) &\equiv \svl{\partial}{b}\svu{\dot{X}}{a}\\&= \Omega^{\mathfrak{ac}} \svl{\partial}{b}\svl{\partial}{c}H.
\end{align}
In other words, $\jmat{a}{b}(t)$ describes how quickly perturbations grow in phase space.  Furthermore, as this matrix is the product of two spatial derivatives, $\jmat{a}{b}(t)$ also encodes information about the curvature of the phase space, which is helpful for building intuition about the stability of $\svu{X}{a}(t)$. Schematically, unstable trajectories will lie on hilltops in appropriate phase space slicings (as depicted by the blue surface in~\fig{fig:vectorplot}).

The evolution of $\delta \svu{X}{a}(t)$ is then
\begin{equation}\label{eq:deltaevolution}
\delta \svu{X}{a}(t) = \lmat{a}{b}(t) \delta \svu{X}{b}(0),
\end{equation}
where
\begin{equation}\label{eq:ldef}
\lmat{a}{b}(t) \equiv \exp(\int_{t_0}^t \jmat{a}{b}(t')dt')
\end{equation}
is known as the \emph{linear stability matrix}~\cite{cornish_chaos_2001,jose_classical_1998}, whose initial condition is $\lmat{a}{b}(0)={\delta^\mathfrak{a}}_{\mathfrak{b}}$. The linear stability matrix controls the evolution of vectors that lie in the space that is tangent to the phase space. As $\delta \svu{X}{a}$ evolves, $\lmat{a}{b}(t)$ will simultaneously coerce it to align with its largest eigenvector~(the orange arrows in \fig{fig:vectorplot}, read left-to-right), while also scaling it exponentially by the corresponding eigenvalue~\cite{strogatz_nonlinear_1994}.  

The \emph{Lyapunov spectrum} is then given by the eigenvalues of $\lmat{a}{b}$. The largest of these eigenvalues (the largest element in the Lyapunov spectrum) is the \emph{principal Lyapunov exponent}, $\lambda$. This quantity can be calculated directly by
\begin{equation}\label{eq:lambdadef}
    \lambda \equiv \lim_{t\rightarrow \infty}\frac{1}{t}\ln\left(\frac{\lmat{a}{a}(t)}{\lmat{b}{b}(0)}\right),
\end{equation}
If $\lambda$ is a positive number, then $\svu{X}{a}(t)$ is an unstable trajectory and the perturbation will grow without bound. If $\lambda$ is an imaginary number, then the trajectory is stable to small perturbations~\cite{cornish_chaos_2001, strogatz_nonlinear_1994, jose_classical_1998}.

It is worth pointing out a few details that will be useful later on.  First, finding Lyapunov exponents analytically is rare. Calculating $\lmat{a}{b}(t)$ requires knowing the path of a full trajectory, $\svu{X}{a}(t)$, which is not, in general, possible for many Hamiltonians.  Second, at no point did we concern ourselves with tracking the evolution of the separation between two initially close trajectories, which is how this subject of Lyapunov exponents is sometimes presented.  This approach can add unnecessary complexity.  Working in the linearized regime prevents this potential pitfall. Third, that the Lyapunov exponent is an eigenvalue suggests that they are well-suited to perturbative calculations, as we show in Appendix~\ref{ap:lyap_pert}. While perturbative techniques are not necessary to complete the work we will present in this paper, it is worth noting the techniques presented here are quite general.  As a result, one could generate Lyapunov exponents for any conceivable modification to the Hamiltonian for null geodesic motion.


\section{Photon Trajectories around Kerr Black Holes}
\label{sec:photonring}

In this section, we briefly review the structure of null trajectories close to a Kerr black hole. Doing so will help establish intuition when we tackle the corresponding problem in modified gravity. For more complete treatments, see~\cite{gralla_lensing_2020,johnson_universal_2020,Gralla_2020,1997}.

\subsection{Geodesic Equations and Analytic Solutions}
In Boyer-Lindquist coordinates, a given trajectory on a Kerr background with Kerr spin parameter $a$ and mass $M$ follows a four-momentum $p^\mu$ that satisfies
\begin{subequations}
\begin{align}
& \frac{\Sigma}{E} p^r= \pm_r \sqrt{\mathcal{R}(r)}, \label{eq:pr}\\
& \frac{\Sigma}{E} p^\theta= \pm_\theta \sqrt{\Theta(\theta)}, \label{eq:pth}\\
& \frac{\Sigma}{E} p^\phi=\frac{a}{\Delta}\left(r^2+a^2-a \ell\right)+\frac{\ell}{\sin ^2 \theta}-a, \label{eq:pph}\\
& \frac{\Sigma}{E} p^t=\frac{r^2+a^2}{\Delta}\left(r^2+a^2-a \ell\right)+a\left(\ell-a \sin ^2 \theta\right), \label{eq:ppt}
\end{align}
\label{eq:evolution}%
\end{subequations}
where $E = -p_t$ and $\ell = p_\phi/E$ are the conserved energy and energy-rescaled angular momentum, respectively, and $\Sigma = r^2+a^2\cos{\theta}$ and $\Delta = r^2 - 2Mr +a^2$ are functions that appear in the metric. The functions $\mathcal{R}(r)$ and $\Theta(\theta)$ are the radial and polar potentials, respectively, which are defined by
\begin{equation}
\begin{aligned}
& \mathcal{R}(r)=\left(r^2+a^2-a \ell\right)^2-\Delta\left[\eta+(\ell-a)^2\right], \\
& \Theta(\theta)=\eta+a^2 \cos ^2 \theta-\ell^2 \cot ^2 \theta,
\end{aligned}
\end{equation}
where $\eta$ is the energy-rescaled Carter constant. The turning points, or maximum and minimum values of $\theta$ of the trajectory's evolution, $\theta_{\pm}$, are given by
\begin{equation}\label{eq:turningpoints}
    \theta_\pm = \arccos\left(\mp\sqrt{u_+}\right),
\end{equation}
where
\begin{equation}
u_{ \pm}=\triangle_\theta \pm \sqrt{\triangle_\theta^2+\frac{\eta}{a^2}}, \quad \triangle_\theta=\frac{1}{2}\left(1-\frac{\eta+\ell^2}{a^2}\right).
\end{equation}

As explored in impressive detail in~\cite{gralla_lensing_2020,Gralla_2020}, these evolution equations permit full analytic solutions in terms of elliptic functions. Of particular use to our current effort, recasting them in integral form will allow the unambiguous definition of a complete orbit of the black hole. We can do this by integrating along the path of a particle traveling from its source point with coordinates $(t_s, r_s, \theta_s, \phi_s)$ to its observed point at $(t_o, r_o, \theta_o, \phi_o)$. While the complete set of integrals is not needed in this paper, it will be useful to cover the $\theta$ case. The curious reader is directed to~\cite{gralla_lensing_2020,Gralla_2020} for the full story.

Let us then define the quantity 
\begin{equation}\label{eq:gthint}
G_\theta=\fint_{\theta_s}^{\theta_o} \frac{\mathrm{d} \theta}{ \pm_\theta \sqrt{\Theta(\theta)}},
\end{equation}
where the slash indicates that the integral should be taken along the trajectory. Equation~\eqref{eq:gthint} can be expressed in terms of elliptic integrals as 
\begin{equation}
G_\theta=\frac{1}{a \sqrt{-u_{-}}}\left[2 m K \pm F_s \mp F_o\right],
\end{equation}
where $m$ is the number of angular turning points encountered in the trajectory,  
\begin{equation}
F_i=F\left(\arcsin \left(\frac{\cos \theta_i}{\sqrt{u_{+}}}\right) \bigg| \frac{u_{+}}{u_{-}}\right)
\end{equation} is the elliptic integral of the first kind (with $i \in \{s,o\}$), and 
\begin{equation}
K = F\left(\frac{\pi}{2}\bigg|\frac{u_{+}}{u_{-}}\right).
\end{equation}

Let us now make a few observations. First, the quantity $G_\theta$, defined in Eq.~\eqref{eq:gthint} is equivalent to the so-called ``Mino time''~\cite{Mino_2003}, which can be used to decouple Eq.~\eqref{eq:evolution}. Second, this notation enables the unambiguous comparison of trajectories that are closed to those that are not by defining one complete orbit to be a complete oscillation in $\theta$.  We therefore here adopt the convention of~\cite{gralla_lensing_2020}, and declare one full orbit to be the traversal from one turning point in $\theta$, as defined in Eq.~\eqref{eq:turningpoints}, back to itself again.

This, coupled with the definition of $G_\theta$ above, allows us to define the number of orbits, $n$, as
\begin{equation}\label{eq:ndef}
    n = \frac{G_\theta}{G^1_\theta},
\end{equation}
where the normalization factor $G^1_\theta$ is the time required to complete one orbit:
\begin{equation}\label{eq:hodef}
G_\theta^1\equiv2 \fint_{\theta_{-}}^{\theta_{+}} \frac{\mathrm{d} \theta}{\sqrt{\Theta(\theta)}}=\frac{4 K}{a \sqrt{-u_{-}}}.
\end{equation}
\begin{figure}
    \centering
\includegraphics[scale=0.575]{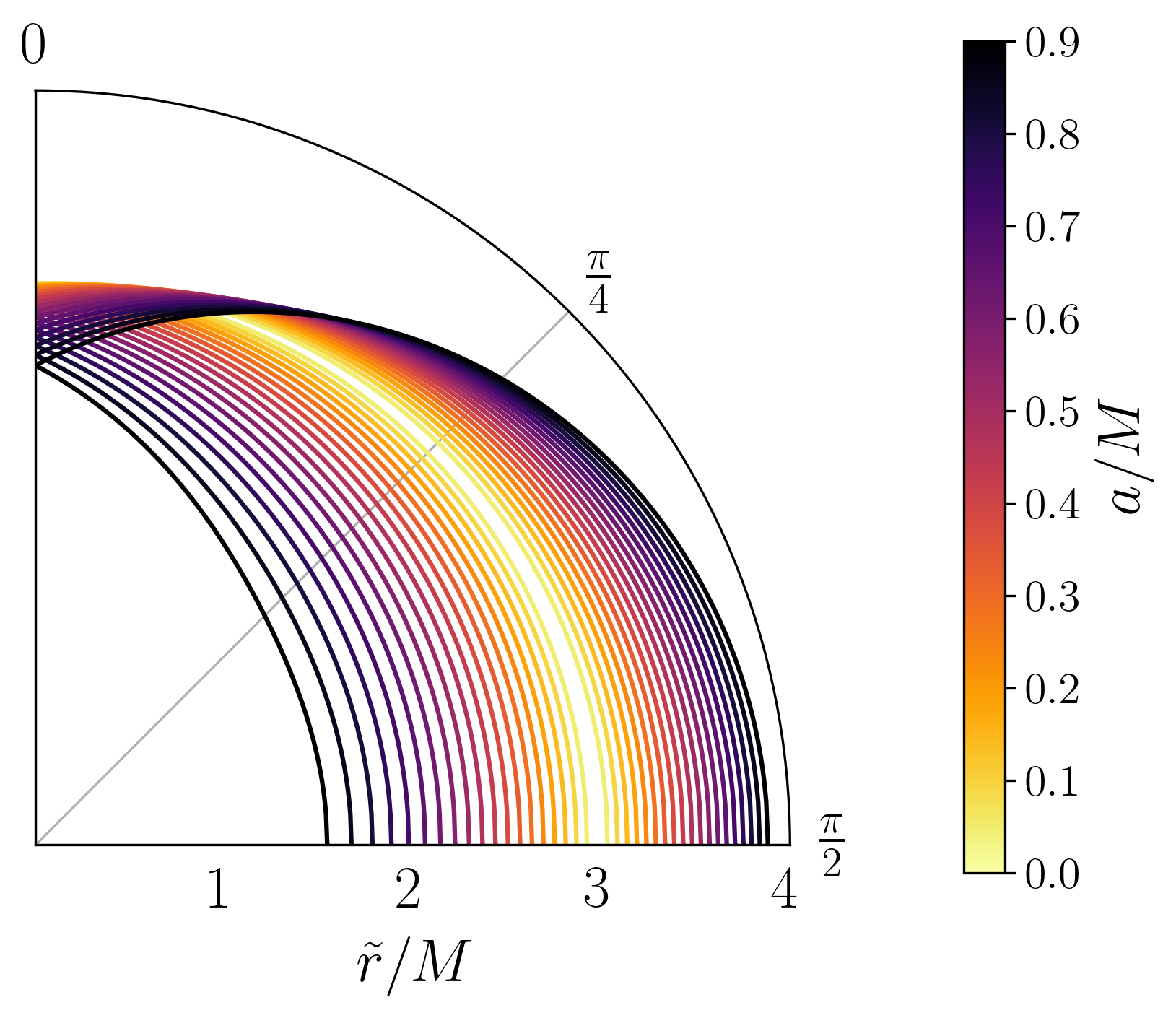}
    \caption{The maximum and minimum values of $\theta$ of a given photon shell orbit are determined entirely by its value of $\tilde{r}$. Pictured here are the values of $\theta_+$ for trajectories around a Kerr black hole across a range of dimensionless spin values. This relationship is axisymmetric, so the $\theta_-$ values are identical to the above, but mirrored along the $\theta = \pi/2$ axis.}
    \label{fig:thvsr0}
\end{figure}


\subsection{The Photon Shell, Rings, and Subrings}

The region near a black hole where there exist trapped null trajectories is called the \emph{photon shell}. In this region, a given trajectory has no radial evolution, but can oscillate in $\theta$ inside $(\theta_-,\theta_+)$, which depends on the chosen value of $r$. For a Kerr black hole, the limiting values of the photon orbit radii are given by
\begin{equation}
    r_{\pm} = 2M\left[1+\cos{\left(\frac{2}{3}\arccos{\left(\pm\frac{a}{M}\right)}\right)}\right],
\end{equation}
where $M$ and $a$ are the mass and spin of the black hole, respectively. Therefore, a photon on the shell has a fixed $r-$coordinate value $\tilde{r}$, with $r_{-} \leq \tilde{r} \leq r_{+}$. If $\tilde{r}=r_{\pm}$, then the trajectory is confined to orbit on the equator. In~\fig{fig:thvsr0}, we show the values of $\theta_-$ and $\theta_+$ as a function of the value of $\tilde{r}$ across a range of spin values for a Kerr black hole.

In the event that a photon trajectory is not exactly bound---i.e. its $r$-coordinate value is close to, but not exactly, the value of $\tilde{r}$ that would place it in a bound orbit in the photon shell---then the photon will either scatter or plunge into the black hole event horizon.  Ultimately, it is these almost-bound photons that are of interest to us, because the bound photons are never seen by observers far from the black hole. These almost-bound photons are the ones responsible for creating an image on the image plane of the observer's screen, which we call the \emph{photon ring}.
\begin{figure}
    \centering
\includegraphics[scale=1.1]{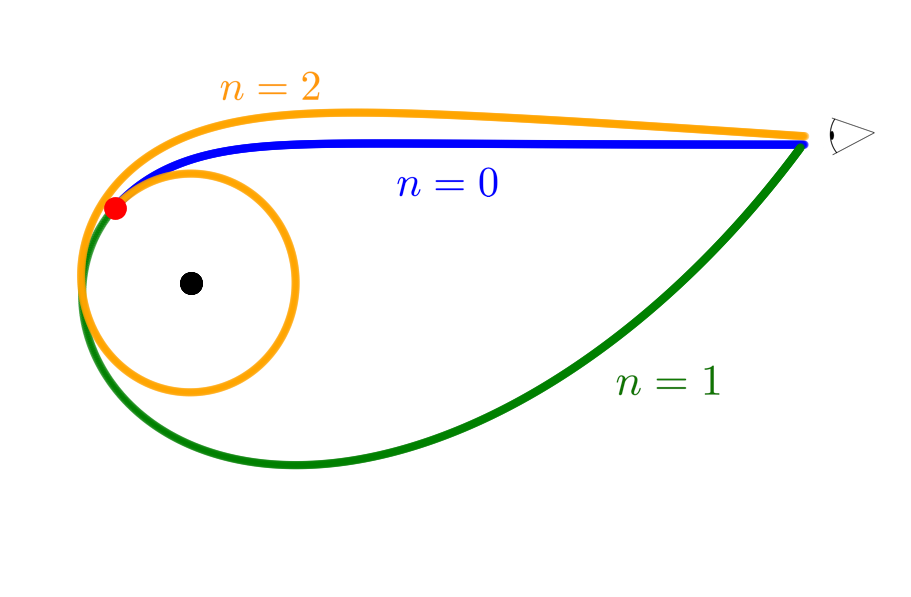}
    \caption{Geodesics connecting some point (red dot) outside the black hole (black dot) to the observer are indexed by the number of half-orbits of the black hole they complete. Pictured are the $n=0,1,2$ equatorial trajectories (blue, green, and orange, respectively) for a Kerr black hole with spin $a = 0.8 M$. Observe how the number of half-orbits can be used to label geodesics.}
    \label{fig:nexample}
\end{figure}
Now, consider what is required of a given set of light rays emanating from some point outside the black hole, if we require that they all eventually impact the detector.  While there are infinitely many such light rays, the different paths they can take are uniquely indexed by the number of half-orbits they complete before hitting the detector, $n$. This is visualized in~\fig{fig:nexample} for the $n=0, 1, 2$ trajectories (the blue, green, and orange lines).  As a result, we can unambiguously refer to a single geodesic that connects the detector to a given point outside the black hole solely by $n$.

\section{Lyapunov Exponents on the Photon Shell}\label{sec:PRlyaps}

Here we cover the specifics of Lyapunov exponents for photon ring trajectories in axisymmetric spacetimes. First, we discuss how Lyapunov exponents are calculated for these geodesics. Then we cover some practical considerations for their observation by comparing the flux ratios from adjacent sub-rings.

\subsection{Calculating Lyapunov Exponents in General Axisymmetric Spacetimes}\label{sec:genlyaps}

\begin{figure*}[htb]
  \centering
  \subfigure{\includegraphics[scale=0.5]{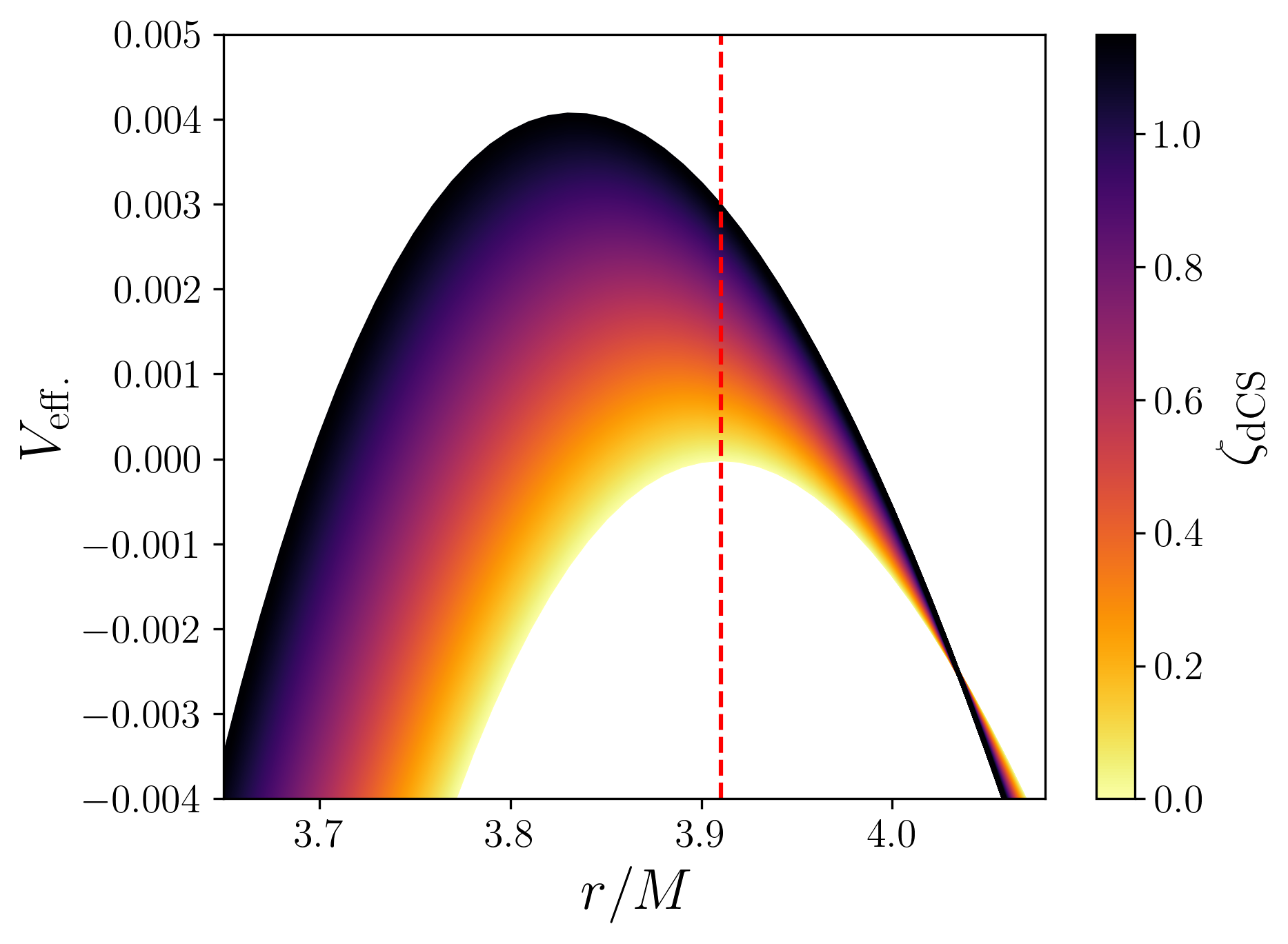}}%
  \quad\subfigure{\includegraphics[scale=0.5]{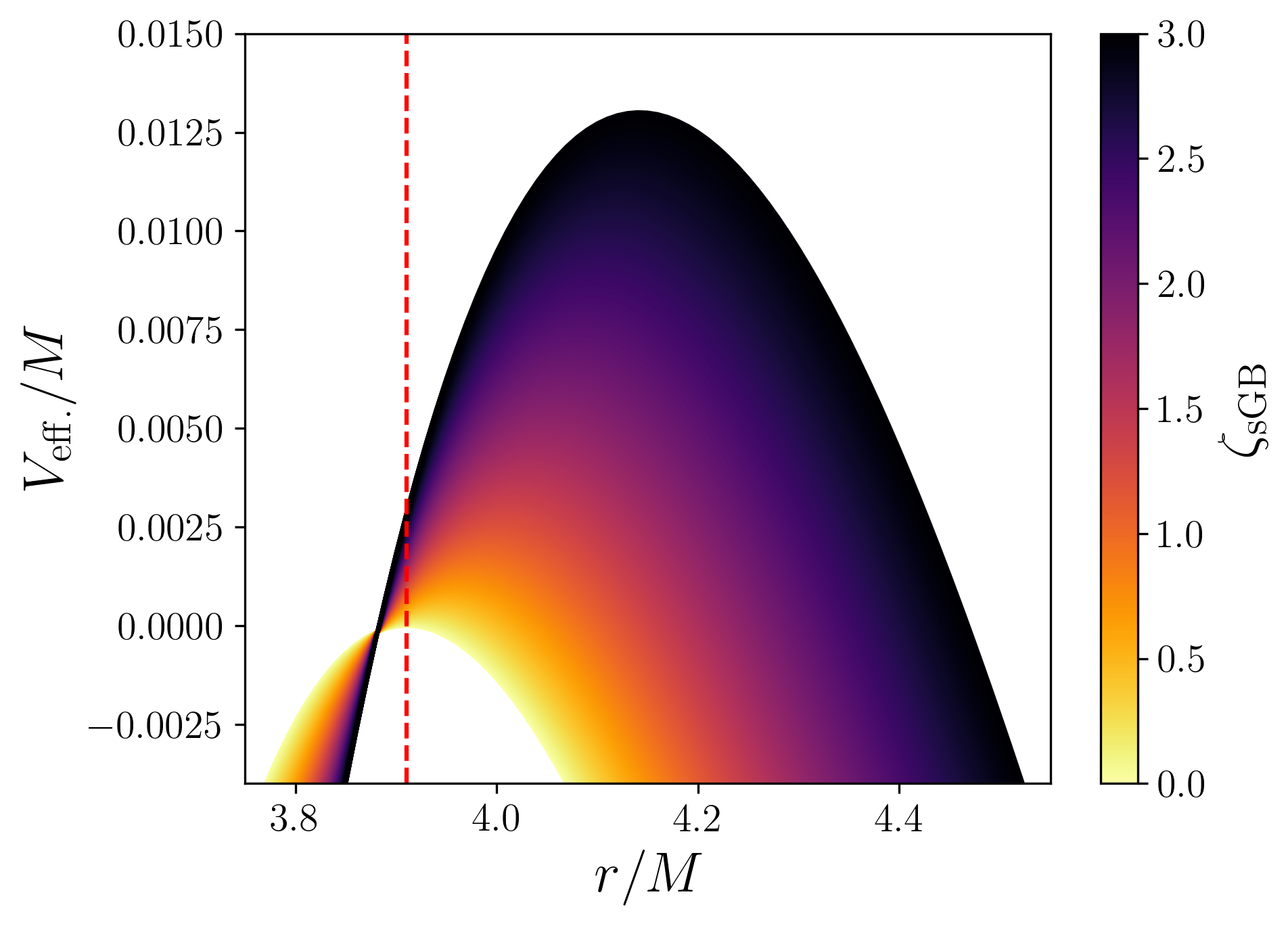}}
    \caption{Null orbit effective potentials in dCS gravity (left) and sGB gravity (right) for a range of values of $\zeta_\text{q}$, with $\chi=0.9$ and $L\approx-6.832$. The photon orbit in Kerr, denoted by the dashed red lines, corresponds to the peak of the $\zeta_{\rm q} = 0$ curve. Notice that the location of this point depends on the coupling parameter. Data from~\cite{cano2023accuracy}.}
    \label{fig:veff_example}
\end{figure*}

Photon shell trajectories are inherently unstable, owing to their position at the top of a local maximum in the effective potential, as shown by the yellow edge in both panels of~\fig{fig:veff_example}. From this figure, we observe that in both dCS (left panel) and sGB (right panel) theories (see Sec.~\ref{subsec:BHsModGR} for their respective black hole metrics), the position of the photon ring shifts with the size of the coupling parameter. Thus, we expect them to possess positive Lyapunov exponents. Here we show one method of analytically calculating these Lyapunov exponents for null geodesics in the general class of axisymmetric spacetimes, to which the Kerr metric belongs. For more details, see~\cite{cornish_lyapunov_2003,johnson_universal_2020}.

In order to simplify the problem, we first restrict ourselves to just those trajectories that are bound to the equator (i.e. $\theta = \pi/2$). We make this simplification for two reasons. First, this restricts the phase space to only 2 dimensions, in $r$ and $p_r$, so that the form of $\jmat{a}{b}$ is as straightforward as possible. Second, these circular geodesics are, for general 
axisymmetric spacetimes, the only geodesics for which closed-form solutions exist. We will exploit the same simplification when we extend this calculation to modified theories of gravity.

The Lagrangian is 
\begin{equation}
    \mathcal{L} = \frac{1}{2}g_{\mu\nu}\frac{dx^\mu}{ds}\frac{dx^\nu}{ds},
\end{equation}
where $g_{\mu\nu}$ is the spacetime metric, $s$ is an affine parameter and $x(s)$ is the trajectory's four-position~\cite{Misner1973}. From this, the canonical conjugate momenta are
\begin{equation}
    p_\mu \equiv \frac{\partial \mathcal{L}}{\partial (dx^\nu/ds)} = g_{\mu\nu}\frac{d x^\mu}{ds}
\end{equation}
Then, performing a Legendre transformation, the Hamiltonian reads
\begin{align}
    H &\equiv p_\mu\frac{dx^\mu}{ds} - \mathcal{L}
    =\frac{1}{2} g^{\mu\nu}p_\mu p_\nu = 0,
\end{align}
where the final equality follows from the fact that we are dealing with null trajectories. Considering only equatorial geodesics, the Hamiltonian above becomes
\begin{equation}\label{eq:simple_ham}
    H = \veff + \frac{p_r^2}{g_{rr}},
\end{equation}
where $\veff$ is the effective potential, given in terms of metric components by
\begin{equation}\label{eq:veffdeff}
    \veff(r) = \frac{L^2 g_{tt}(r)+E^2g_{\phi\phi}(r)+2 E L g_{t\phi}(r)}{g_{tt}(r)g_{\phi\phi}(r)-g_{t\phi}(r)^2},
\end{equation}
where $E$ and $L$ are the conserved energy and angular momenta of the trajectory, respectively. In the absence of analytic expressions for the trajectories, the location and angular momentum of the equatorial photon orbit is found by setting $\theta = \pi/2$ and solving the system
\begin{equation}\label{eq:PRsystem}
    \veff = 0 = \partial_r \veff
\end{equation}
simultaneously for $\tilde{r}$ and $L$. Doing so will usually result in two real solutions, corresponding to $r_-$ and $r_+$.

Then, setting $p_r=0$, as all photon shell trajectories require, the evolution Jacobian~\eqref{eq:deltadef}, with proper time parameterization, takes the form
\begin{equation}
    \jmat{a}{b} = \begin{pmatrix}
    0 & -g_{rr}^{-1} \\
\partial_r^2\veff & 0
\end{pmatrix},
\end{equation}
whose eigenvalues, evaluated at $r=\tilde{r}$, are the Lyapunov exponents
\begin{equation}\label{eq:lyapsqrt}
    \lambda_{\rm p} = \pm \sqrt{-\frac{\partial_r^2\veff}{g_{rr}}}\Bigg\rvert_{r=\tilde{r}},
\end{equation}
which we note is dependent upon the full set of spacetime parameters, including spin.

Let us pause here for a moment to make clear an important point. Lyapunov exponents are not invariant under changes of the time parametrization used. Rather, it is the ratio between the Lyapunov timescale $\tau_\lambda = 1/\lambda_{\rm p}$ and a relevant timescale that is invariant and may be compared between systems. In other words, the numerical value of a given trajectory's Lyapunov exponent will change if calculated in, say, proper time parametrization versus Schwarzschild time. But if we also calculate a timescale in the same parametrization, $\tau'$, then the ratio $\tau'/\tau_\lambda = \lambda_{\rm p} \tau'$ will be invariant under reparametrization. In the case currently being considered, the relevant timescale is the time required for a half-orbit, $G^1_\theta$, as given by Eq.~\eqref{eq:hodef}. Therefore, it is not sufficient to simply calculate Eq.~\eqref{eq:lyapsqrt}; we must also calculate the time required for one complete orbit.

\subsection{Detectability of Sub-Rings and the Measurement of the Lyapunov Exponent by Future VLBIs}
\label{subsec:detectability}

By tracking the fractional number of orbits completed by a trajectory as it evolves, we have actually created for ourselves another affine coordinate with which we can parametrize our trajectories. This is a handy parametrization for calculations involving the detection of the photon ring, as the fractional number of orbits determines the specific sub-ring to which the photon belongs. Thought of this way, we can consider how a perturbation in $r$ evolves over some number $n$ of orbits,
\begin{equation}\label{eq:fracorblyap}
    \delta r(n) = \exp(\lambda_{\rm M}n)\delta r(0),
\end{equation}
where $\lambda_{\rm M}$ is the principal Lyapunov exponent parametrized by the fractional number of orbits. 

In general then, let us define $\lambda_{\rm M}$ as follows
\begin{equation}
    \lambda_{\rm{M}} = \lambda_{\rm{p}} \tau_{\rm{M}}\,,
\end{equation}
where $\lambda_{\rm p}$ is calculated via Eq.~\eqref{eq:lyapsqrt} and $\tau_{\rm M}$ corresponds to the time it takes to complete a half-orbit for a null geodesic around a rotating black hole. For a Kerr black hole, this time is simply given in closed form by $\tau_{\rm M,Kerr} = G_\theta^1$ in Eq.~\eqref{eq:gthint}. In this case then, one can show~\cite{johnson_universal_2020} that $\lambda_{\rm M, Kerr}$ takes the analytic form
\begin{align}\label{eq:kerrLEdef}
\lambda_{\rm M,Kerr} &=  \sqrt{\frac{\partial_r^2\veff}{2g_{rr}}} G_\theta^1\Bigg\rvert_{r=\tilde{r}} = \frac{4 \tilde{r} \sqrt{\tilde{\chi}}}{a \sqrt{-\tilde{u}_{-}}} \tilde{K},
\end{align}
where
\begin{equation}
\tilde{\chi} \equiv1-\frac{M \Delta(\tilde{r})}{\tilde{r}(\tilde{r}-M)^2}
\end{equation}
and where each quantity with a tilde is understood to be evaluated on the photon shell. When defined like this, the principal Lyapunov exponent parameterized by the fractional number of orbits, $\tau_{\rm M}$, implicitly gains the factor of $G^{1}_\theta$ mentioned earlier. 

As a trajectory starts closer and closer to an exactly bound orbit, and $n$ gets high, the image on the detector screen approaches a closed curve known as the \emph{critical curve}~\cite{johnson_universal_2020}. Giving this curve dimensionless detector screen angular coordinates $(\rho_c, \phi_c)$, it can then be shown that any photons that impact at $\delta \rho$ near $\rho_c$ must be funneled into an exponentially narrowing annulus of width
\begin{equation}\label{eq:rholambd}
    \frac{\delta \rho}{\rho_c} \approx \exp(\lambda_{\rm M} n).
\end{equation}
This means that each sub-ring is sequentially nested according to the number of half-orbits completed en route. When one integrates Eq.~\eqref{eq:rholambd} over a solid angle to determine the flux generated by each sub-ring, one finds~\cite{gralla_lensing_2020,johnson_universal_2020}
\begin{equation}\label{eq:srflux}
    \frac{F^{n+1}}{F^n}\approx \exp(-\lambda_{\rm M})
\end{equation}
for the ratio in flux recieved between adjacent subrings.

\section{Calculating Lyapunov Exponents in Modified Gravity}\label{sec:modgrav}

Here we introduce the field equations of two varieties of modified gravity theories, dCS and sGB gravity, and discuss the rotating black hole solutions in each. This allows us to then explain our method of calculating the half-orbit timescale and, consequently, the Lyapunov exponents of photon ring null orbits in these theories.

\subsection{Rotating Black Holes in sGB and dCS gravity}
\label{subsec:BHsModGR}

There are two main ways of motivating quadratic gravity.  First, if we assume the Einstein-Hilbert (EH) action is simply the leading-order term in a more general effective field theory, we can modify the standard EH action by adding terms that are expansions in curvature. Second, these quadratic theories also occur naturally from low-energy expansions of certain string theories\cite{alexanderyunes2009,dcs2003}. These theories have an action that reads
\begin{equation}\label{eq:genaction}
    S = S_{\text{EH}} + S_{\text{mat}} + S_\vt + S_{RR},
\end{equation}
where $S_\text{EH}$ is the EH action, $S_\text{mat}$ is the matter action, $S_\vt$ is an action for a dynamical scalar or pseudo-scalar field, and $S_{RR}$ couples a quadratic curvature term to the field. The only distinction between the two quadratic theories we are concerned with is in this final term. The EH action reads
\begin{equation}
    S_\text{EH} = \kappa \int d^4x \, \sqrt{-g} \, R \,,
\end{equation}
where $R=g^{\alpha\beta}g^{\rho\sigma}R_{\rho\alpha\sigma\beta}$ the Ricci scalar, and $\kappa = (16\pi)^{-1}$, $g$ is the determinant of the metric tensor with the Riemann tensor $R_{\rho\alpha\sigma\beta}$. Meanwhile, the scalar or pseudo-scalar field action is
\begin{equation}
\label{eq:action-sf}
    S_\vt = -\frac{1}{2}\int d^4x \sqrt{-g} \left[\nabla_\mu\vt\nabla^\mu\vt + \nabla_\mu\vt\nabla^\mu\vt + 2V(\vt)\right],
\end{equation}
where the potential of the scalar field is $V(\vt)$.  In order to preserve the shift symmetry of $\vt \rightarrow \vt + \text{const.}$, we set $V(\vt)=0$ to specify a massless theory, which is a feature often found in effective string theories at low energy, including in both dCS and sGB gravity~\cite{PhysRevD.93.029902}.

For sGB and dCS gravities specifically, we can prescribe a curvature-coupling action that generically encompasses both theories. Let us then define
\begin{multline}\label{eq:genquad}
S_{RR} = \int d^4x\sqrt{\left|g \right|} \left\{\alpha_{\rm sGB} \vt_{\rm sGB} RR + \alpha_{\rm dCS}\vt_{\rm dCS} R\tilde{R} \right\},
\end{multline}
where
\begin{equation}
RR = R_{\mu\nu\rho\sigma}R^{\mu\nu\sigma}-4R_{\mu\nu}R^{\mu\nu}+R^2    
\end{equation}
 is the so-called Gauss-Bonnet density, and where
 \begin{equation}
    R \tilde{R} \equiv {}^*R^\alpha{}_{\beta}{}{}^{\gamma\delta}R^{\beta}{} _{\alpha\gamma\delta}\,,
\end{equation}
is the Pontryagin density with ${}^*R^{\alpha}{}_{\beta}{}^{\gamma\delta} = \frac{1}{2}\epsilon^{\gamma\delta\rho\lambda}R^{\alpha}{}_{\beta\rho\lambda}$ the dual of the Riemann tensor. With this in hand, sGB gravity is defined by setting $\alpha_{\rm dCS} = 0$ in Eq.~\eqref{eq:genquad} and $\vartheta = \vartheta_{\rm sGB}$ in Eq.~\eqref{eq:action-sf}, while dCS gravity is defined by setting $\alpha_{\rm sGB}=0$ in Eq.~\eqref{eq:genquad} and $\vartheta = \vartheta_{\rm dCS}$ in Eq.~\eqref{eq:action-sf}. The parameters $\alpha_{\rm sGB}$ and $\alpha_{\rm dCS}$ determine the coupling parameter strength of the particular theory being described, and they have dimensions of length squared in geometric units. In this paper, we report results after non-dimensionalizing the coupling parameter of either theory, so we define
\begin{equation}
    \zeta_{\rm q} \equiv \frac{\alpha_{\rm q}^2}{\kappa M^4},
\end{equation}
where $\rm q$ is $\rm dCS$ or $\rm sGB$, depending on the theory in question. We choose the maximum values of $\zeta_{\rm q}$ that we will explore by finding the maximum values that still satisfies the small-coupling approximation, i.e.~the maximum value of $\zeta_{\rm q}$ that generates a correction to the metric components that remains small relative to the Kerr metric components everywhere outside the black hole. Explicitly, we demand that $|\zeta_{\rm q} H_i| < 0.5$, where the $H_i$ are the functions that perturb the metric components (see e.g.~\cite{cano2023accuracy}). Doing so results in maximum values of $\zeta_{\rm dCS}^{\rm max}  = 1.15$ and $\zeta_{\rm sGB}^{\rm max} = 12.5$.

\begin{figure*}[htb]
  \centering
  \subfigure{\includegraphics[scale=0.56]{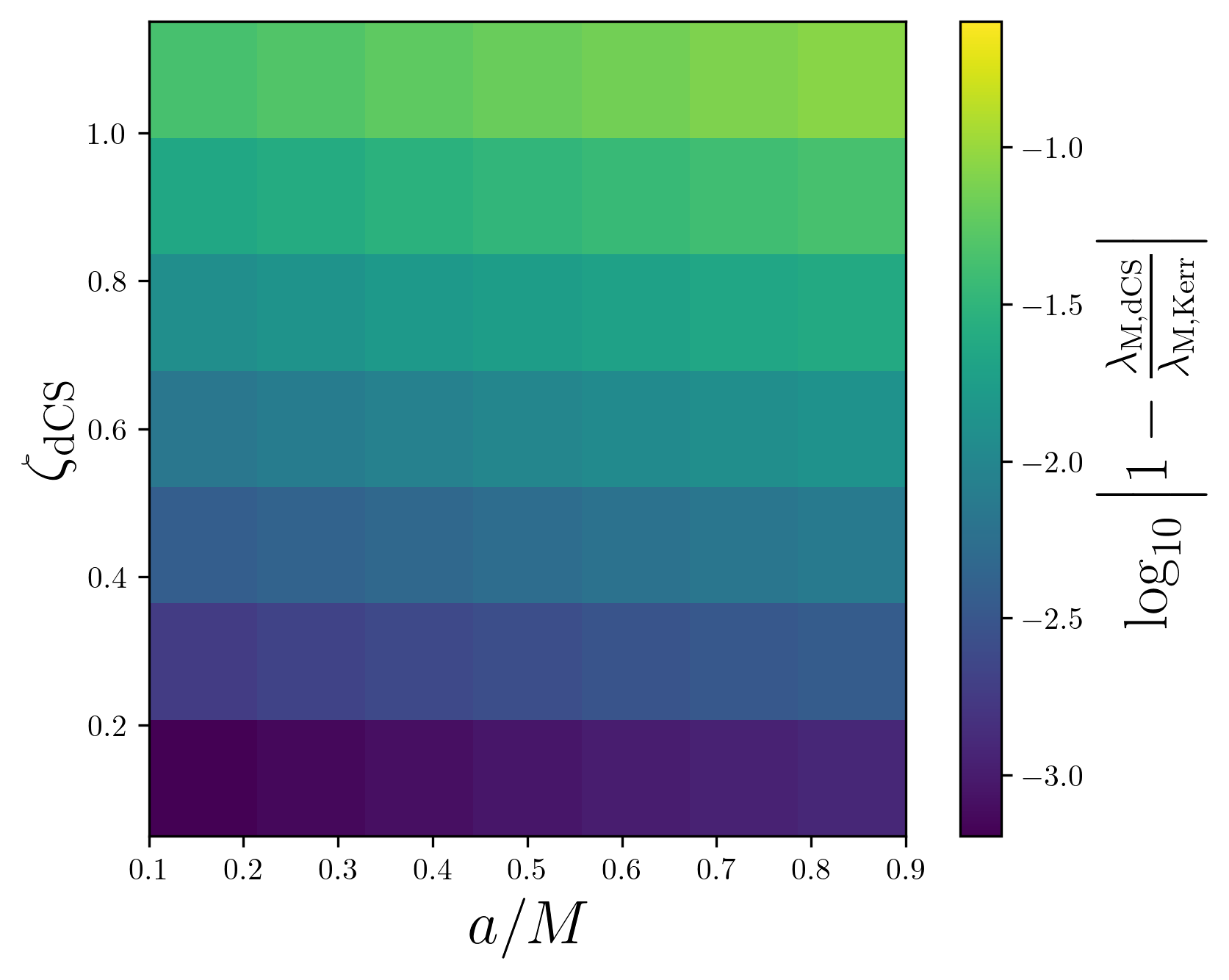}}%
    \quad\subfigure{\includegraphics[scale=0.56]{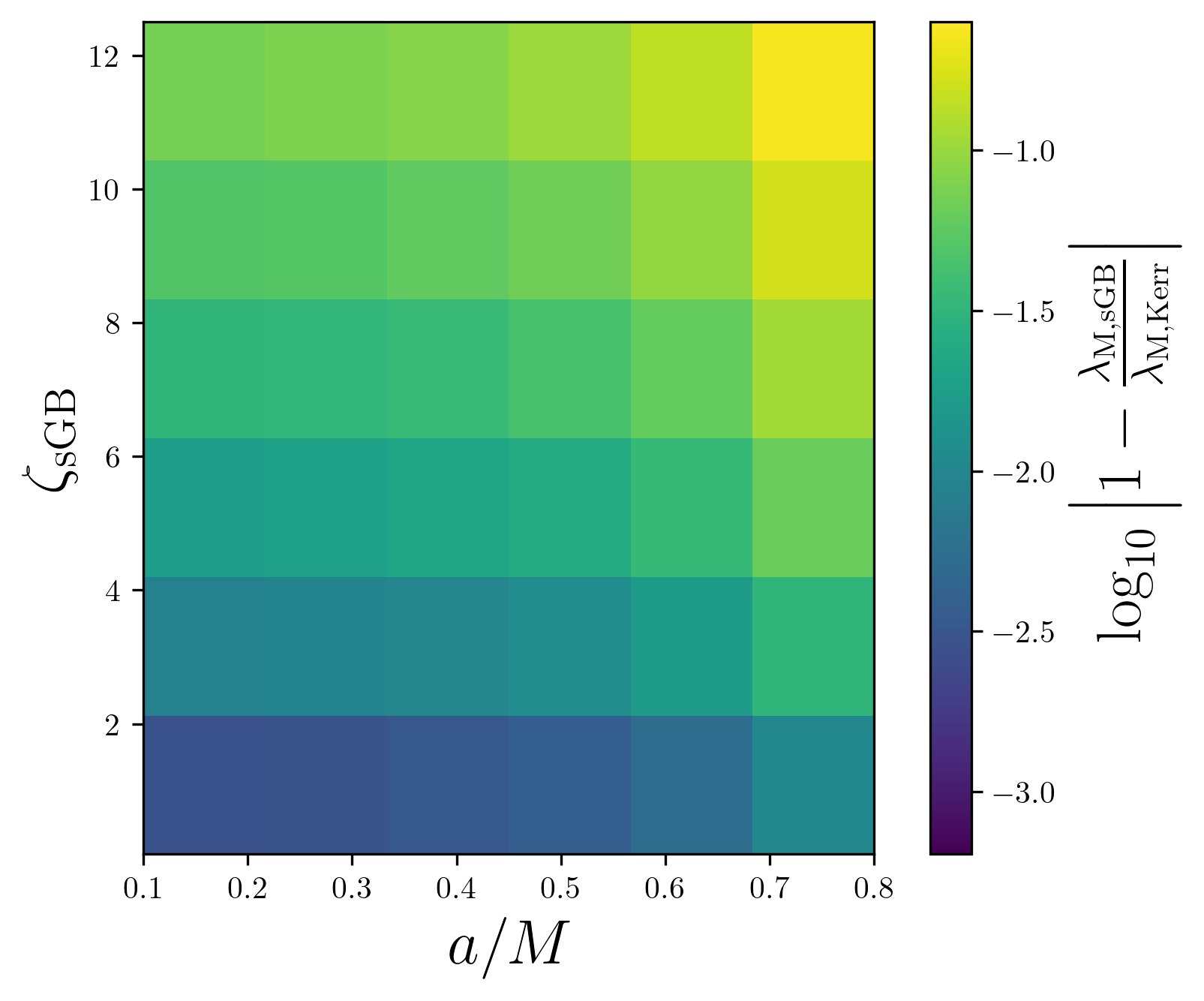}}
    \caption{The log fractional difference in the calculated Lyapunov exponent for equatorial null geodesics around dCS (left) and sGB (right) black holes across a grid of dimensionless spin and coupling parameter.}
    \label{fig:result}
\end{figure*}

Our two example theories have different motivations and phenomenology, and, therefore, different constraints.  SGB theory is motivated by a certain low-energy limit of string theory~\cite{Kanti:1995vq}. Unique among these two theories, sGB gravity induces modifications in the spacetime regardless of whether the spacetime is spherically symmetric (i.e., regardless of whether the black hole is spinning or not), and it induces dipole scalar radiation in black hole binaries. This is why gravitational wave observations have already constrained $\alpha_{\rm sGB}^{1/2} \leq 5.6 \; \text{km}$ at 90\% confidence~\cite{constraints1}. 

On the other hand, dCS gravity finds motivation from a few sources, including loop quantum gravity~\cite{dcs2003,yagi_challenging_2016}, the standard model gravitational anomaly~\cite{dcs2003,Yagi:2013mbt}, and investigations in string theory\footnote{A combination of both theories, sGB and dCS, with two scalar fields, can also arise in the effective action of heterotic string theory \cite{Cano:2021rey} if, somehow, the dilaton is not stabilized.}~\cite{Campbell:1990fu,yagi_slowly_2012}.  Unlike the sGB case, dCS gravity does not modify spherically symmetric spacetimes, and thus non-rotating black hole, and it does not activate dipole scalar radiation in binaries. For this reason, dCS gravity has not yet been constrained with gravitational waves alone, but rather, the most stringent bounds come from neutron star multi-messenger observations, which lead to $\alpha_{\rm dCS}^{1/2} \leq 8.5\text{km}$ within a 90\% confidence interval~\cite{constraints2}.  

The field equations of the theories described above are so complicated that analytic exact solutions that represent black holes with arbitrary rotation have not yet been found. Instead, we are forced to work with metrics that are simultaneous expansions in small coupling and dimensionless spin, $\chi\equiv a/M$, where $a = S/m$ and $S$ is the (magnitude of the) black hole spin angular momentum. While the expansion in the coupling $\zeta_{\rm q}$ is always kept to linear order (because these theories are effective), recent work~\cite{Cano_2019,Cano:2021rey} has made it possible to achieve arbitrary orders in the spin expansion. To label these expansions, we adopt the notation $\mathcal{O}(\zeta_{\rm q} \chi^m)$ for a metric that is expanded to $m^{\rm th}$ order in dimensionless spin. Written in a generic form, the metric components are expanded to  ${\cal{O}}(\zeta_{\rm z} \chi^m)$ in the following way:
\begin{equation}
g_{\mu \nu}=g_{\mu \nu}^{\mathrm{Kerr}}+\zeta^{\prime} \sum_{\ell=0}^{\ell=m} \left(\chi^{\prime}\right)^{\ell} \delta g_{\mu \nu}^{(\ell)},
\end{equation}
where $\zeta^{\prime}$ and $\chi^{\prime}$ are bookkeeping parameters labeling the expansion in coupling and expansion in spin, respectively, and $g^{\rm Kerr}_{\mu \nu}$ is the Kerr metric.
For the work presented here, we find that the quantities of interest achieve sufficient accuracy at $\mathcal{O}(\zeta_{\rm q} \chi^{14})$ for both dCS and sGB theories.  Finally, we note that the metrics available for sGB are not valid for $\chi > 0.8$~\cite{cano2023accuracy}, which is the upper-bound for spin in the work presented here.

Finally, we note that if the constraints from gravitational wave and neutron star observations are saturated, then $\zeta_{\rm q}$, for supermassive black holes, takes values significantly smaller than the maximum values we consider here (if we take Sgr A*, then we find $\zeta_{\rm sGB}=1.2 \times 10^{-24}$ and $\zeta_{\rm dCS}=2.8 \times 10^{-24}$, respectively). However, these values for $\alpha^{1/2}_{\rm q}$ were derived from objects with masses smaller by many orders of magnitude than the supermassive black holes which would be the targets of any VLBI observation~\cite{Johnson_2023,constraints2}.  Further, the method we present here is, in principle, an independent test.  Therefore, any deviations that are found could motivate a higher-order EFT which screens larger modifications at small scales.

\subsection{Generalized Equatorial Half-Orbit Timescales}
We can apply the machinery developed in Sec.~\ref{sec:genlyaps} to also calculate $\tau$ for equatorial photon orbits in generalized spacetimes, such as those provided by sGB and dCS. To solve for $\tau$, we still look for eigenvalue solutions to Eq.~(\ref{eq:deltadef}) as before, but with two key differences. First, we restrict our phase space to motion in $\theta$ only.  Second, in order to retrieve the harmonic solutions, we demand that the eigenvalues be imaginary. Thus, this eigenvalue, labeled $\omega_\theta$, represents the frequency of oscillations in $\theta$.

When this is carried through, one finds for $\omega_\theta$ an expression very similar to that found for $\lambda_{\rm p}$ earlier (Eq.~\ref{eq:lyapsqrt}):
\begin{equation}
    \omega_{\theta} = \sqrt{\frac{\partial_\theta^2\veff}{g_{\theta \theta}}}\Bigg\rvert_{r=\tilde{r}}.
\end{equation}
Then, finding the half-orbit timescale via $\tau = \pi/\omega_\theta$, we have all we need to calculate $\lambda_{\rm p}$ in arbitrary spacetimes. All together, our expression for $\lambda_{\rm M}$ reads
\begin{equation}
    \lambda_{\rm M} = \pi\sqrt{-\frac{g_{\theta\theta}}{g_{rr}}\frac{\partial_r^2\veff}{\partial_\theta^2\veff}}\Bigg\rvert_{r=\tilde{r}}.
\end{equation}
We were able to check the validity of this method by developing a numerical approach, which is described in Appendix~\ref{ap:numerical_timescales}.  This approach calculates $\tau$ for arbitrary values of $\theta$, and yields answers within accuracy of $10^{-6}$.

With this in hand, we can now compute the principal Lyapunov exponent parameterized in terms of half orbits through Eq.~\eqref{eq:lambdadef} and compare this to its Kerr value in Eq.~\eqref{eq:kerrLEdef}. When we do this, we find the largest fractional difference to be of ${\cal{O}}(10^{-1})$ for geodesics around dCS black holes with dimensionless spin of $a/M = 0.9$ and coupling parameter $\zeta_{\rm dCS} = 1.15$ (left panel of~\fig{fig:result}), and ${\cal{O}}(10^{-0.5})$ for geodesics around sGB black holes with dimensionless spin of $a/M = 0.8$ and coupling parameter $\zeta_{\rm sGB} = 12.5$ (right panel of~\fig{fig:result}).

These results allow us to make several observations. First, dCS and sGB modifications to Lyapunov exponents increase with the coupling constant and spin. Therefore, the best target of future VLBI observations to constrain these theories would be rapidly spinning black holes. Both of these theories, however, are already constrained by other astrophysical observations (see Sec.~\ref{subsec:BHsModGR}), and thus, the value of the coupling constant cannot be increased without bound.

Notice that the results presented here should not be understood as necessarily making a direct claim about the detectability of modified theories with this method. Such a statement would depend on, among other factors, an accurate and independent measurement of the BH dimensionless spin.  This is because the modified gravity correction to $\lambda_{\rm p}$ depends both on $\zeta_{\rm q}$ and $a/M$.  In the absence of this data, our results should instead be read as a lower-bound on the accuracy of the BH dimensionless spin measurement required before entertaining constraining a modified theory.

\section{Conclusions}
\label{sec:conclusions}

We have here constructed a framework to calculate Lyapunov exponents in a theory-agnostic way that allows the direct computation of the flux ratio between adjacent sub-rings in VLBI images.  We then applied this framework to two theories of modified gravity, dCS and sGB gravity, and calculated the log fractional difference between these theories and the Lyapunov exponent in a Kerr spacetime.  We find the corrections are of ${\cal{O}}(10^{-1})$ for geodesics around dCS black holes with dimensionless spin of $a/M = 0.9$, and ${\cal{O}}(10^{-0.5})$ for geodesics around sGB black holes with dimensionless spin of $a/M = 0.8$.

However, our results do not necessarily imply that you can constrain either theory, due to a number of confounding factors present in real-world observations. First, a measurement of $\lambda_{\rm M}$ would need to be made to within the accuracy of the log fractional differences we present. This would require, among other things, that the uncertainties due to the astrophysical environment be smaller than any (most probably very small) deviation from GR. Second, any statement about constraints on the theory would require disentangling the effect on $\lambda_{\rm M}$ due to $\zeta_{\rm q}$ from that due to the black hole spin, most probably necessitating an independent measurement of the latter quantity.

We expect these results to be of interest in two ways. One, they inform our knowledge of the behavior of photons very near black holes, and will be of interest when designing observating campaigns of these environments. Second, the procedure developed here has applications in theory more broadly. The work we present is very general, being applicable to any conceivable modification to the metric or even the Hamiltonian itself, especially given the prospect of being able to calculate these effects perturbatively.

In the future, there are at least two avenues that would extend this work.  One is to continue hunting for a means of measuring the differences in Lyapunov exponents due to quadratic gravity theories. An approach that exploits the auto-correlations of the photon ring~\cite{Hadar_2021, Chen_2023} could possibly amplify the effect, perhaps rendering more detectable the small differences calculated here. Another route would be to explore different terms that one could add to the Hamiltonian to change the size and shape of the effective potential. Such terms could include, for instance, the presence of a third body, or yet further modifications to the theory of gravity itself.

\acknowledgements

We would like to thank Alex Lupsasca for several discussions, including the suggestion to use auto-correlations in future work. We also would like to thank Leo Stein for discussions about symplectic structure.  
AD and NY acknowledge support from the Simmons Foundation through Award No.~896696, and the NSF through awards PHY-2207650 and WoU-2007936.

\appendix
\section{Perturbation of Local Lyapunov Exponents in Curved Backgrounds}\label{ap:lyap_pert}

The perturbation of eigenvalues of symplectic matrices follows closely the standard story from linear algebra. The only point of possible confusion is that there is no well-defined means of ``raising'' an index of a symplectic object as one may be accustomed to in differential geometry. Instead, upper- and lower-indexed eigenvectors are simply those vectors that solve the right- and left-eigenvalue problems, respectively. That is to say, for a symplectic matrix $\jmat{a}{b}$, vectors $\svu{R}{a}$ and $\svl{L}{b}$ are said to be its eigenvectors if they satisfy
\begin{align}
    \svl{L}{a}\jmat{a}{b} = \lambda \svl{L}{b}\label{eq:Ldef},\\
    \jmat{a}{b}\svu{R}{b} = \lambda \svl{R}{a}\label{eq:Rdef},
\end{align}
where $\lambda$ is the corresponding eigenvalue. Now, considering the following series expansions, 
\begin{align}
    \jmat{a}{b} &= \jmat{\textnormal{(0)}a}{b} + \alpha \jmat{\textnormal{(1)}a}{b}\label{eq:jpert} + ...,\\
    \svu{R}{a} &= \svu{R}{\textnormal{(0)}a} + \alpha\svu{R}{\textnormal{(1)}a}\label{eq:Rpert} + ...,\\
    \svl{L}{a} &= \svl{L}{\textnormal{(0)}a} + \alpha\svl{L}{\textnormal{(1)}a} + ...,\\
    \lambda &= \lambda^{(0)} + \alpha \lambda^{(1)}+...,
\end{align} where the parenthetical exponents denote expansion order and $\alpha$ is a bookkeeping parameter. We can now follow through with the standard eigenvalue perturbation. We start by right-multiplying Eq.~\eqref{eq:jpert} by Eq.~\eqref{eq:Rpert} to find
\begin{align}
    &\left(\jmat{\textnormal{(0)}a}{b} + \alpha \jmat{\textnormal{(1)}a}{b} + ...\right)\left(\svu{R}{\textnormal{(0)}b} + \alpha\svu{R}{\textnormal{(1)}b} + ...\right) \\=
    &\left(\lambda^{(0)} + \alpha\lambda^{(1)} + ... \right)\left(\svu{R}{\textnormal{(0)}a} + \alpha\svu{R}{\textnormal{(1)}a} + ...\right)\,.
\end{align}
Then, keeping only the linear-order terms, we find
\begin{align}\label{eq:step3}
    \jmat{\textnormal{(1)}a}{b}\svu{R}{\textnormal{(0)}b} + \jmat{\textnormal{(0)}a}{b}\svu{R}{\textnormal{(1)}b} = \lambda^{(1)}\svu{R}{\textnormal{(0)}a}+\lambda^{(0)}\svu{R}{\textnormal{(1)}a}.
\end{align}
Now, we can left-multiply both sides of Eq.~\eqref{eq:step3} by $\svl{L}{\textnormal{(0)}a}$ to find
\begin{align}
    \svl{L}{\textnormal{(0)}a}\jmat{\textnormal{(1)}a}{b}\svu{R}{\textnormal{(0)}b} + \svl{L}{\textnormal{(0)}a}\jmat{\textnormal{(0)}a}{b}\svu{R}{\textnormal{(1)}b} = \\
    \lambda^{(1)}\svl{L}{\textnormal{(0)}a}\svu{R}{\textnormal{(0)}a}+\lambda^{(0)}\svl{L}{\textnormal{(0)}a}\svu{R}{\textnormal{(1)}a},
\end{align}
and use Eq.~\eqref{eq:Ldef} and \eqref{eq:Rdef} to cancel the second term on the left-hand side with the second term on the right-hand side, leaving us with
\begin{align}
    \svl{L}{\textnormal{(0)}a} \; \jmat{\textnormal{(1)}a}{b} \; \svu{R}{\textnormal{(0)}b}=\lambda^{(1)},
\end{align}
after normalization, $\svl{L}{\textnormal{(0)}a}\svu{R}{\textnormal{(0)}a} = 1$ (so we drop that term from the final expression). The above expression gives us the familiar result that the first-order perturbation to the eigenvalue is given by contracting the unperturbed eigenvectors onto the first-order matrix perturbation. The fact that this machinery translates perfectly to the symplectic context means that Lyapunov exponents can be calculated for any imaginable perturbation to the Hamiltonian. 

\section{Numerically Calculated Half-Orbit Timescales}\label{ap:numerical_timescales}
We develop here a method to calculate half-orbit timescales numerically to high precision. First, we use the geodesic evolution equations in second-order form, which after a trivial first-order reduction are 
\begin{equation}\label{eq:evolutioneqs}
    \frac{dx^\mu}{ds} = \frac{\partial H}{\partial p_\mu}, \qquad \frac{dp_\mu}{ds} = -\frac{\partial H}{\partial x^\mu},
\end{equation}
Now, we must follow what is suggested by Eq.~\eqref{eq:ndef} and calculate the time required for a single half-orbit. In order to do so, one option would be to simply track a trajectory's numerical evolution in $\theta$, starting at $\theta_+$, and solve numerically for the proper time it takes for the trajectory to reach $\theta_-$. However, this method is needlessly computationally intensive, as it would require using something like a bisection algorithm to accurately determine when the trajectory has crossed $\theta_-$.

A better route is to instead invert the $\dot{\theta}$ evolution equation and integrate between the angular turning points to find the proper time interval we are after. To reduce the accumulated error further, we can exploit the system's axisymmetry and integrate only to the equator, multiplying by 2 to find the half-orbit proper time interval. For a given null trajectory with four-velocity $dx^{\mu}/ds$ in Boyer-Lindquist-like coordinates, we may then define
\begin{equation}\label{eq:taumdef}
    \tau_{\rm M,q} \equiv  2 \int^{\pi/2}_{\theta_+} \frac{ds}{d\theta}d\theta
\end{equation} 
as the time required to complete a half-orbit\footnote{This is the same as the ``Mino time'' for the trajectory, hence the subscript. For more on the Mino time, see~\cite{gralla_lensing_2020,Gralla_2020,johnson_universal_2020}.}.The problem has then reduced to finding $d\theta/ds$ as a function of $\theta$. We therefore use the chain rule to rewrite the relevant Hamilton equations as

\begin{align}
    \frac{d x^a}{d\theta} &= \frac{d x^a}{ds} \frac{d s}{d\theta} = \left(\frac{\partial H}{\partial p_a} \right) \left(\frac{\partial H}{\partial p_\theta}\right)^{-1}, 
   \nonumber \\
   \frac{d p_c}{d\theta} &=  \frac{d p_c}{ds}  \frac{d s}{d\theta} = -\left(\frac{\partial H}{\partial x^c}\right)
   \left(\frac{\partial H}{\partial p_\theta}\right)^{-1},
\end{align}
where $a\in[r,\phi]$ and $c\in[r, \theta]$, so that we can integrate the full, coupled system simultaneously with respect to $\theta$.

Before we initialize these trajectories, we must deal with an issue that occurs because we are restricting ourselves to equatorial orbits, which ostensibly experience no evolution in $\theta$ (or in other words, $\theta_+ = \pi/2$, and so the definition of $\tau_{\rm M,q}$ in Eq.~\eqref{eq:taumdef} would be meaningless).  We can deal with this by finding $\tau_{\rm M,q}$ for a series of trajectories whose initial $\theta(0)$ value is shifted from the equator by an amount $\delta \theta$, and tracking the value that $\tau_{\rm M,q}$ approaches as $\delta \theta \rightarrow 0$, before using a Richardson extrapolation~\cite{doi:10.1098/rsta.1911.0009} to calculate the final value.

The above method also determines the order in which the evolution equations are initialized. Setting $\theta=\pi/2+\delta\theta$ as described above, we can demand that the trajectory begins at the top of its $\theta$ trajectory by setting $p_\theta = 0$ (this therefore sets $\theta_+ = \pi/2 + \delta\theta$ and $\theta_- = \pi/2 - \delta\theta$). 
Then, we can again solve the system in Eq.~\eqref{eq:PRsystem}(evaluated at the aforementioned $\theta$) simultaneously to find the initial values of $r$ and $L$. This allows us to finally integrate the equations in~\eqref{eq:taumdef} and~\eqref{eq:evolutioneqs} from $\theta_+$ to $\pi/2$, and thereby extract a value for $\tau_{\rm M,q}$ through Eq.~\eqref{eq:taumdef}. We used a Runge-Kutta integrator of order 8 due to Dormand and Price~\cite{hairer2008solving}, with relative tolerance set to $10^{-13}$ and absolute tolerance set to $10^{-14}$. We found that using three trajectories with $\delta \theta = 10^{-6}, 10^{-7},$ and $10^{-8}$ was sufficient to achieve an accuracy in $\tau_{\rm M}$ of $10^{-6}$. This was verified by computing $\tau_{\rm M,Kerr}$ numerically (i.e.~$\tau_{\rm M}$ for null geodesics around a Kerr black hole) and comparing the result to the analytically exact value of Eq.~\eqref{eq:gthint}.

\bibliographystyle{apsrev4-1}
\bibliography{bib.bib}
\end{document}